\documentclass[a4paper]{llncs}
\usepackage[english]{babel}
\usepackage[utf8]{inputenc}
\usepackage{amssymb,amsmath}
\usepackage{url}
\usepackage{float}
\usepackage{graphicx}
\usepackage{multirow,array}
\usepackage{booktabs}

\begin{document}


\title{On Solving the Maximum $k$-club Problem}

\author{Andreas Wotzlaw} 

\institute{Institut für Informatik, Universität zu Köln,
  Weyertal 121, D-50931 Köln, Germany\\
  \email{wotzlaw@informatik.uni-koeln.de}}

\maketitle

\begin{abstract}
  Given a simple undirected graph $G$, the maximum $k$-club problem is
  to find a maximum-cardinality subset of nodes inducing a subgraph of
  diameter at most $k$ in $G$. This NP-hard generalization of clique,
  originally introduced to model low diameter clusters in social
  networks, is of interest in network-based data mining and clustering
  applications. We give two MAX-SAT formulations of the problem and show
  that two exact methods resulting from our encodings outperform
  significantly the state-of-the-art exact methods when evaluated both
  on sparse and dense random graphs as well as on diverse real-life
  graphs from the literature.

\end{abstract}

\section{Introduction}
Let $G=(N,E)$ be a simple undirected graph with the set $N$ of $n$ nodes
and the set $E$ of $m$ edges. The length of a shortest path between two
nodes $i$ and $j$ in $G$ is denoted by $dist_G(i,j)$, whereas
$d_G:=\max_{i,j\in N}dist_G(i,j)$ is the {\em diameter} of $G$. For a
nonempty subset of nodes $S\subseteq N$, $G[S]$ denotes the subgraph
$(S,E(S))$ of $G$ induced by $S$ on $G$, where $E(S)$ are edges of $E$
with both end nodes in $S$. If every pair of nodes $i,j\in S$ is
connected in $G[S]$ by at least one path with at most $k$ edges, in
other words, $d_{G[S]}$ is at most $k$, then $S$ is called a {\em
  $k$-club} of $G$.

The {\em maximum $k$-club problem}, M$k$CP, consists in finding a
maximum cardinality $k$-club in $G$. We denote the cardinality of a
maximum $k$-club in $G$ by $\omega_k(G)$, referred to also as the {\em
  $k$-club number} of $G$. A $k$-club is regarded as diameter-based
relaxation of {\em clique}~\cite{veremyev2012identifying}. Recall, a
clique $C$ in $G$ is a subset of $N$ such that the subgraph $G[C]$ of
$G$ is complete, i.e., $d_{G[C]}=1$. Hence, for $k=1$, the definition of
$k$-club is equivalent to that of clique. 

The notion of $k$-club was introduced in social network
analysis~\cite{mokken1979cliques} as an alternative way to model tightly
linked groups of actors (e.g., people, companies, web communities or
sites), referred to as {\em cohesive
  subgroups}~\cite{scott2000social}. In those groups every member is
related to all other members either directly or via other
members. Although cliques are useful for modeling high-density
communities~\cite{bomze1999maximum}, they appear to be too restrictive
to represent real-life groups where rarely all members are connected
directly. Here, the idea of $k$-club can be used instead to model
low-diameter clusters in graphs. It finds its application in graph-based
data mining in social, biological, financial, and communication
networks~\cite{almeida2012integer,balasundaram2005novel,shahinpour2013algorithms}.

\paragraph{Related work.} M$k$CP is computationally
challenging. Bourjolly et al.~\cite{bourjolly2002exact} established the
NP-hardness of M$k$CP, even for fixed $k>1$, and proposed an
\mbox{exact} branch-and-bound algorithm for it. Balasundaram et
al.~\cite{balasundaram2005novel} showed that M$k$CP remains NP-hard even
when restricted to graphs of fixed diameter. Unlike cliques, the
$k$-club model is of nonhereditary nature~\cite{mokken1979cliques},
meaning that every subset of a $k$-club is not necessarily a $k$-club
itself. An important manifestation of this property is the
intractability of testing maximality of $k$-clubs, demonstrated by
Mahdavi Pajouh and Balasundaram~\cite{pajouh2012oninclusionwise}. They
also developed a branch-and-bound technique $\mathsf{B\&B}$ for M$k$CP
using the $k$-coloring number as an upper bound. For fixed $k\geq 2$,
Asahiro et al.~\cite{asahiro2010approx} proved that M$k$CP is
inapproximable within a factor of $n^{\frac{1}{2}-\epsilon}$ for any
$\epsilon > 0$, unless P=NP. M$k$CP is fixed-parameter tractable when
parameterized by solution size as shown by Schäfer et
al.~\cite{schaefer2012param}.
In~\cite{hartung2012parametrized,hartung2013onstructural}, Hartung et
al. gave recently a systematic classification of the complexity of M2CP
with respect to several structural graph parameters like, e.g., feedback
edge set size, as well as a new well-performing parameterized algorithm
for M2CP. Moreover, Schäfer~\cite{schaefer2009exact} demonstrated that
M$2$CP on bipartite graphs can be solved in $O(n^5)$ time, whereas
M$k$CP on trees and interval graphs needs $O(nk^2)$ and $O(n^2)$ time,
respectively. Chang et al.~\cite{chang2013finding} proved recently that
M$k$CP can be solved exactly in $O^*(1.62^n)$ time, where $O^*$ hides
factors polynomial in $n$. The first polyhedral results for $2$-club
polytope were given in~\cite{balasundaram2005novel}. While M$k$CP has a
compact Boolean integer programming (BIP) formulation for $k=2$, the
formulations proposed for $k\geq 3$
in~\cite{balasundaram2005novel,carvalho2011upper} need exponentially
many variables. Alternative BIP formulations for $k=3$ were explored by
Almeida and Carvahlo~\cite{almeida2012integer}. The first
polynomial-size BIP formulation for a general $k$ using $O(kn^2)$
variables and constraints was given by Veremyev and
Boginski~\cite{veremyev2012identifying}. Further, Chang et
al.~\cite{chang2013finding} implemented a branch-and-bound algorithm for
M$k$CP using a new heuristic IDROP for finding initial lower
bounds. Finally, Shahinpour and Butenko~\cite{shahinpour2013algorithms}
presented a well-performing exact branch-and-bound method for M$k$CP
using variable neighborhood search for lower bounding.

\paragraph{Our contribution.} In this paper, we present a new exact
approach for M$k$CP. To this end, we give first in Section~\ref{s:model}
two propositional-logic-based formulations of M$k$CP. In both cases, we
encode M$k$CP on graph $G$ as an instance of the PARTIAL MAX-SAT
problem~\cite{cha1997local} of some propositional formula in conjunctive
normal form with a mandatory part of clauses that must be satisfied for
the solution to be reasonable, and a second part of clauses of length 1
(1-clauses), such that a truth assignment must satisfy as many of them
as possible. In an optimal solution of such a PARTIAL MAX-SAT instance,
the number of satisfied 1-clauses is than equal to $\omega_k(G)$, and
the Boolean variables assigned by the truth assignment to $1$ indicate
the nodes of $G$ included in an optimal $k$-club of $G$. Our first
satisfiability-based formulation of M$k$CP needs $O(n^{k-1})$ variables
and clauses, whereas for the second one $O(kn^2)$ variables and
$O(kn^3)$ clauses suffice.

According to the experimental evaluation for $k\in \{2,3,4\}$ (i.e., for
typical values of $k$ from the
literature~\cite{almeida2012integer,balasundaram2005novel,shahinpour2013algorithms})
given in Section~\ref{s:evaluation}, our exact methods $\mathsf{SatMC1}$
and $\mathsf{SatMC2}$ for M$k$CP incorporating the encodings from
Section~\ref{s:model}, when compared with a straightforward exact
BIP-based approach using the problem formulation described
in~\cite{almeida2012integer,veremyev2012identifying}, as well as with
two well-performing specialized exact branch-and-bound methods
$\mathsf{VNS}$~\cite{shahinpour2013algorithms} and
$\mathsf{B\&B}$~\cite{pajouh2012oninclusionwise}, demonstrate clearly
their practical strength by outperforming the other three methods
considerably. Also, they offer a simple yet effective alternative for
finding good-quality approximate solutions for M$k$CP, as the numerical
results for our both methods show. Finally, in
Section~\ref{s:conclusion} we conclude our work and state some open
questions.

\section{Satisfiability-based Formulation of M$k$CP}
\label{s:model}
\paragraph{Preliminaries.} Let CNF denote the set of propositional
formulas in conjunctive normal form over a set $V$ of Boolean
variables. Each variable $x\in V$ induces a positive literal (variable
$x$), or a negative literal (negated variable $\overline{x}$). Each
formula $C\in$ CNF is regarded as a {\em set} of its clauses. Similarly,
a clause is considered as a {\em set} of its literals. A clause is
termed a {\em $k$-clause}, for some integer $k>0$, if it contains
exactly $k$ literals. We denote by $V(C)$ the set of variables occurring
in formula $C$. The satisfiability problem (SAT) asks whether formula
$C$ is~\emph{satisfiable}, i.e., whether there is a truth assignment $t
: V(C) \rightarrow \{0, 1\}$ setting at least one literal in each clause
of $C$ to 1, whereas for every $x\in V$ it holds
$t(x)=1-t(\overline{x})$. Given a formula $C \in$ CNF, the optimization
version MAX-SAT searches for a truth assignment $t$ satisfying as many
clauses of $C$ as possible, whereas in its PARTIAL variant some clauses
(called {\em hard}) must be satisfied.

\paragraph{Our Method.} We only consider simple undirected graphs
$G=(N,E)$ with $N =\{1,...,n\}$ and $m:=|E|$. Each node is referred to
by its number. Let $A:=(a_{ij})$ be the adjacency matrix of $G$, where
the values $a_{ij}$'s are regarded as constant truth values 0 and 1,
such that $a_{ij}=1$ iff an edge $\{i,j\}\in E$, for $1\leq i,j \leq n$.

We are now ready to give our two PARTIAL MAX-SAT formulations of M$k$CP
on graph $G$ for an integer $k>1$. For this purpose, we define for every
node $i\in N$ a Boolean variable $x_i$, such that $x_i=1$ if and only if
$i$ belongs to a specific $k$-club of $G$. We proceed next in two
steps. We define first a CNF formula $C_S$ ensuring the optimality,
i.e., the maximum cardinality, of a solution $S\subseteq N$ to M$k$CP on
$G$. In the second step, we show the construction of two CNF formulas,
$C_{H}$ and $D_{H}$, for the first and the second formulation,
respectively, both consisting only of hard clauses and ensuring the
correctness of a solution $S$ to M$k$CP, i.e., $S$ is a $k$-club in
$G$. The unions $C_S\cup C_{H}$ and $C_S\cup D_{H}$ will give finally
the first and the second PARTIAL MAX-SAT encoding of M$k$CP on $G$,
respectively.

The formula $C_S$ consists of 1-clauses solely and is defined as
follows:
$$
C_S := \{\{x_1\}, ..., \{x_n\}\}.
$$

Now we construct $C_{H}$ for the first encoding as follows:
$$
C_{H} := \bigcup_{i=1}^{n-1}\bigcup_{j\in \{i+1,...,n\}|a_{ij}=0}
\{C_{ij}\},\quad\mbox{ where }\quad 
C_{ij} := \{\overline{x}_i,\overline{x}_j\}\cup \bigcup_{l=1}^{k-1}
C^l_{ij}
$$ 
and
$$
C^l_{ij} := 
\bigcup_{r_1\in N_*}\bigcup_{r_2\in N_1}...\bigcup_{r_l\in N_{l-1}}
\{x_{r_1}\wedge x_{r_2} \wedge ...\wedge x_{r_l}\;|\; a_{ir_1}\wedge
a_{r_1r_2} \wedge ... \wedge a_{r_lj}=1\},
$$
where $N_*:=N\setminus\{i,j\}$ and $N_p:=N_*\setminus \{r_1,...,r_p\}$,
for $p=1,...,l-1$. Note, that each conjunction $x_{r_1}\wedge x_{r_2}
\wedge ...\wedge x_{r_l}$ in $C_{ij}^l$ together with nodes $i,j$
corresponds to a path of length $l+1$ from $i$ to $j$ in $G$. Due to
$N_*$ and $N_p$ in the definition of $C_{ij}^l$, no paths with cycles
can be generated, what may result in a tighter encoding. However, for
the correctness of $C_H$, these restrictions of $N$ are not necessary.

To finish our construction, $C^l_{ij}$, for $l\geq 2$, has to be
transformed into a clause. For this, we replace each occurrence of
$x_{r_1}\wedge x_{r_2} \wedge ...\wedge x_{r_l}$ in $C^l_{ij}$, for all
$i,j\in N$, with a new Boolean variable $y_{r_1...r_l}$ and define $l+1$
additional clauses
$$
\{\overline{y}_{r_1...r_l},x_{r_1}\},\{\overline{y}_{r_1...r_l},x_{r_2}\},
...,\{\overline{y}_{r_1...r_l},x_{r_l}\}, \{\overline{x}_{r_1},
\overline{x}_{r_2},..., \overline{x}_{r_l}, y_{r_1...r_l}\},
$$
expressing after some elementary transformations the logical equivalence
$$
x_{r_1}\wedge x_{r_2} \wedge ...\wedge x_{r_l} \leftrightarrow
y_{r_1...r_l}.
$$
Clearly, for $k=2$, we need $O(n)$ variables and $O(n^{2})$
clauses. However, for $k>2$, $C_{H}$ requires, in consequence of the
transformation of $C_{ij}^l$ into a clause, $O(n^{k-1})$ variables and
clauses. Note, that $C_{ij}$ is generated only if $a_{ij}=0$.

Let $t$ be a truth assignment satisfying $C_H$, and $S$ the nodes
selected by $t$, i.e., $S=\{i\in N\,|\,t(x_i)=1\}$. For the correctness
of $C_{H}$, it suffices to show which conditions do hold in $G[S]$, if a
pair of distinct nodes $i,j\in N$ belongs to $S$, implying $t(x_i)=
t(x_j)=1$. Obviously, only the case $a_{ij}=0$ need to be considered. In
that case, $C_{ij}$ can be satisfied if and only if there exists at
least one conjunction $x_{r_1} \wedge ...\wedge x_{r_l}$ in $C_{ij}^l$
(or, equivalently, at least one variable $y_{r_1...r_l}$ together with
the corresponding clauses after the transformation of $C^l_{ij}$ in a
clause given above), for some $l\in\{1,...,k-1\}$, satisfied by $t$,
i.e., $t(x_{r_1})=...=t(x_{r_l})=1$, implying that nodes $r_1,...,r_l$
belong to $S$, too. Consequently, the nodes $i$ and $j$ are connected in
$G[S]$ via $l$ many nodes $r_1,...,r_l$ on a path from $i$ to $j$ in
$G[S]$. Thus, $d_{G[S]}\leq k$ holds for $S$ specified by $t$ and, by
the definition of $k$-club, we conclude that $S$ is a $k$-club in $G$ if
and only if $t$ satisfies $C_H$.

Finally, observe that solving M$k$CP on $G$ is equivalent in terms of
propositional calculus to determining a truth assignment satisfying
$C_{H}$ and maximizing the number $\tau$ of satisfied clauses in
$C_S$. Clearly, $\tau$ corresponds to $\omega_k(G)$, thus completing the
description of our first satisfiability-based formulation of
M$k$CP. Note that this formulation works trivially for $k=1$.
\begin{theorem}
  Let $G$ be a simple undirected graph, $k$ some positive integer, and
  $t: V(C_{S}\cup C_{H})\rightarrow \{0,1\}$ a truth assignment
  satisfying $C_{H}$ and maximizing the number $\tau$ of satisfied
  clauses in $C_S$. Then $\tau$ is equal to $\omega_k(G)$. Moreover, for
  $k>2$, $C_{S}\cup C_{H}$ contains $O(n^{k-1})$ Boolean variables and
  clauses.
\end{theorem}

The formulation above is in the worst case of an exponential size and
requires explicit enumeration of all paths of length at most $k$ between
all pairs of nodes. Nevertheless, for typical values of
$k$~\cite{almeida2012integer,balasundaram2005novel,shahinpour2013algorithms},
$C_{H}$ is of reasonable size. For $k=2$, we need only $n$ variables and
$n(n+1)/2-m$ clauses (mostly 2-clauses for sparse graphs). For $k=3$, at
most $n+m$ variables and $n(n+1)/2+2m$ clauses suffice.

For $k>3$, our second formulation of M$k$CP, $C_S\cup D_{H}$, is
substantially smaller than the first one, as we shall show it now by
constructing the CNF formula $D_{H}$. For this, we introduce for every
pair of distinct nodes $i,j\in N$ and $l=2,...,k$ a new Boolean variable
$v_{ij}^l$, such that $v_{ij}^l=1$ if and only if there exists at least
one path of length at most $l$ from node $i$ to node $j$ in the subgraph
$G[S]$ induced by the nodes of a $k$-club $S$ of $G$. Initially, for
$l=2$, we can write
$$
v_{ij}^2 \leftrightarrow x_i\wedge x_j \wedge \left(\bigvee_{r=1}^n
  a_{ir}\wedge a_{rj} \wedge x_r\right),
$$
what after some elementary transformations is equivalent to the CNF
formula
$$
D_{ij}^2\hspace{-2pt}:=\hspace{-2pt}\left\{
\{\overline{v}_{ij}^2,x_i\},
\{\overline{v}_{ij}^2,x_j\},
\{\overline{v}_{ij}^2, x_{r_1}, ..., x_{r_p}\},
\{\overline{x}_i,\overline{x}_j,v_{ij}^2,\overline{x}_{r_1}\},...,
\{\overline{x}_i,\overline{x}_j,v_{ij}^2,\overline{x}_{r_p}\}
\right\}
$$
where the nodes $r_1,...,r_p\in\{r\in N\,|\, a_{ir}\wedge
a_{rj}=1\}$. If no such a node exists, then we set $D_{ij}^2:= \{
\{ \overline{v}_{ij}^2\}\}$.

For $l\geq 3$, $v_{ij}^l$ can be defined recursively as
$$
v_{ij}^l \leftrightarrow x_i \wedge \left(\bigvee_{r=1}^{n}
  a_{ir}\wedge v_{rj}^{l-1}\right),
$$
what again after some transformations is equivalent to the CNF formula
$$
D_{ij}^l:=\left\{
\{\overline{v}_{ij}^l,x_i\},\{\overline{v}_{ij}^l, v^{l-1}_{r_1j}, ..., 
v^{l-1}_{r_pj}\},
\{\overline{x}_i,v_{ij}^l,\overline{v}_{r_1j}^{l-1}\},...,
\{\overline{x}_i,v_{ij}^l,\overline{v}_{r_pj}^{l-1}\}
\right\},
$$
where $r_1,...,r_p\in \{r\in N\,|\,a_{ir}=1\}$. If no such a node
exists, then $D_{ij}^l := \{ \{ \overline{v}_{ij}^l\}\}$.

Finally, we define
$$
D_{H}:= \bigcup_{i\in N}\bigcup_{j\in N\setminus \{i\}|a_{ij}=0} D_{ij},
\quad \mbox{ where }\quad
D_{ij}:= \left\{\{\overline{x}_i,\overline{x}_j,
  v_{ij}^2,...,v_{ij}^k\}\right\} \cup \bigcup_{l=2}^kD_{ij}^l.
$$

Observe first that $D_{ij}$ has to be generated only if
$a_{ij}=0$. Moreover, to encode $D_{H}$ for $k>1$, we need $O((k-1)n^2)$
variables and $O((k-1)n^3)$ clauses. Thus, the encoding size remains
polynomial in the input size. Now, similarly to $C_{ij}$, for every pair
of distinct nodes $i,j\in N$, the existence of a satisfying truth
assignment $t$ for $D_{ij}$ with $t(x_i)=t(x_j)=1$ implies that
$dist_{G[S]}(i,j)\leq k$ for $S\subseteq N$ specified by $t$. Hence, $S$
is a $k$-club in $G$ if and only if $t$ satisfies $D_H$.

Finally, note that solving M$k$CP on $G$ is equivalent to determining a
truth assignment satisfying $D_{H}$ and maximizing the number of
satisfied clauses in $C_S$, completing the description of our second
PARTIAL MAX-SAT formulation of M$k$CP. Obviously, this formulation works
fine also for $k=1$.
\begin{theorem}
  Let $G$ be a simple undirected graph, $k$ some positive integer, and
  $t: V(C_{S}\cup D_{H})\rightarrow \{0,1\}$ a truth assignment
  satisfying $D_{H}$ and maximizing the number $\tau$ of satisfied
  clauses in $C_S$. Then $\tau$ is equal to $\omega_k(G)$. Moreover, for
  $k>1$, $C_{S}\cup D_{H}$ contains $O(kn^2)$ Boolean variables and
  $O(kn^3)$ clauses.
\end{theorem}

\section{Comparative Evaluation}
\label{s:evaluation}
\paragraph{Experimental Setup.} The goal of our experiments was to
evaluate, for typical values of $k\in \{2,3,4\}$ from the
literature~\cite{almeida2012integer,balasundaram2005novel,shahinpour2013algorithms},
the performance of two exact methods for M$k$CP, $\mathsf{SatMC1}$ and
$\mathsf{SatMC2}$, implemented in C++ according to the first and the
second encoding from Section~\ref{s:model}, respectively. We tested our
methods against a BIP-based approach $\mathsf{IPMC}$ using M$k$CP
formulations from~\cite{almeida2012integer,veremyev2012identifying}, and
two state-of-the-art exact methods: the hybrid algorithm for M$k$CP
from~\cite{shahinpour2013algorithms}, denoted here by $\mathsf{VNS}$,
and the branch-and-bound technique
$\mathsf{B\&B}$~\cite{pajouh2012oninclusionwise}. To make the study
better comparable with the previous results, we use the same C++
implementations of $\mathsf{VNS}$ and $\mathsf{B\&B}$ as the ones being
tested in~\cite{shahinpour2013algorithms}.

For solving the PARTIAL MAX-SAT instances produced by $\mathsf{SatMC1}$
and $\mathsf{SatMC2}$, we applied a complete MAX-SAT solver clasp
2.1.3~\cite{gebser2007conflict}, an example of a modern SAT solver. It
extends the backtrack search procedure DPLL~\cite{Handbook}, commonly
used for SAT-solving, with efficient conflict-driven clause learning
(CDCL), lazy data structures, deletion polices for learned clauses, and
periodical restarts of the search procedure, among others. For more
details on the key techniques of DPLL- and CDCL-based SAT-solving, we
refer to~\cite{Handbook}. In $\mathsf{IPMC}$, for solving BIP-instances
we use CPLEX 12.1~\cite{cplex121}. All tests were run on a machine with
Intel Xeon E5410 2.33 GHz processor running a 64-bit Linux 3.2.51 with
32GB RAM. All programs (solvers) were run with default call parameters
in a single-threaded mode with only one CPU core permitted. 

There were two sets of graph instances being tested. The first set
contained 12 connected simple graphs from the 10th DIMACS Implementation
Challenge~\cite{dimacs2012}. We used them to test the methods on some
real-life networks ranging from small and dense ones to large and sparse
ones (see Table~\ref{table:res_dimacs_stat}). The graphs of the second
set were generated randomly by the algorithm proposed by Gendreau et
al.~\cite{gendreau1993solving}. This
\begin{table}
  \centering
  \caption{Statistics on DIMACS instances. Here, $n$ and $m$ give the number of
    nodes and edges, $d$ the edge density, and $\omega_2$, $\omega_3$, 
    $\omega_4$ the club numbers computed by $\mathsf{SatMC\{1,2\}}$.}
  \begin{tabular}{lrrrrrr}
    \toprule
    \multirow{1}{*}{Instance} & 
    \multirow{1}{27pt}{\centering $n$} & 
    \multirow{1}{31pt}{\centering $m$} &  
    \multirow{1}{31pt}{\centering $d$} &  
    \multirow{1}{23pt}{\centering $\omega_2$} &
    \multirow{1}{23pt}{\centering $\omega_3$}  &
    \multirow{1}{23pt}{\centering $\omega_4$}  \\
    \midrule
    adjnoun & 112 & 425 & 0.0684 & 50& 82& 107\\
    football & 115 & 613 & 0.0935 & 16& 58& 115\\
    jazz & 198 & 2742 & 0.1406 & 103& 174& 192\\
    celegansm & 453 & 2025 & 0.0198 & 238& 371& 432\\
    email & 1133 & 5451 & 0.0085 & 72& 212& 651\\
    polblogs & 1490 & 16715&0.0151&352&776& 1127\\
    add20 & 2395 & 7462 & 0.0022 & 124& 671&1454\\
    data & 2851 & 15093 & 0.0037 & 18& 32&52 \\
    3elt & 4720 & 13722 & 0.0012 & 10& 16&27 \\
    add32 & 4960 & 9462 & 0.0008 &32 & 99&268\\ 
    hep-th & 8361 & 15751 &0.0006&51&120&344\\
    whitaker3&9800&28989&0.0006 & 9&15&23\\ 
    \bottomrule
  \end{tabular}
  \label{table:res_dimacs_stat}
\end{table}
generalization of the classical uniform random graph generator has
earlier been used for testing new methods for
M$k$CP~\cite{almeida2012integer,bourjolly2002exact,pajouh2012oninclusionwise,shahinpour2013algorithms}.
The edge density of the graphs produced by this method was controlled by
two parameters $a$ and $b$ ($0\leq a \leq b \leq 1$). The {\em expected}
edge density $D$ is $(a+b)/2$, and the node degree variance (NDV)
increases with the increase in $b-a$. In our tests, we used {\em
  connected} graphs with $n=100, 150,$ and 200 and $D=0.035, 0.05, 0.1,
0.15$, and $0.2$, i.e., from the range of challenging instances
according to~\cite{almeida2012integer,bourjolly2002exact}. For each
graph size $n$ and density $D$, we generated 10 samples with minimum NDV
($a=b=D$) and 10 samples with maximum NDV ($a=0$, $b=2D$), denoted in
the following by min and max, respectively. As
in~\cite{almeida2012integer} and in contrast
to~\cite{pajouh2012oninclusionwise,shahinpour2013algorithms}, we decided
to reject samples with more than one component since for them the number
of nodes would be misleading. It would correspond mostly to the
cardinality of the largest component. Though, the maximum $k$-club may
or may not be located in the largest or the most dense component as
indicated already in~\cite{pajouh2012oninclusionwise}.

The running time limit for solving each instance tested was set to 3600
seconds for each method. If an instance could not be solved into
optimality within that time, the computation has been terminated, the
best solution computed so far (i.e., a lower bound for the optimum), as
well as the upper bound have been recorded, and an optimality gap,
(upper bound $-$ best solution size)/(upper bound), has been
reported. The CNF formulas generated by our methods for the DIMACS
instances included up to 3.5 millions variables and 50 millions
clauses. $\mathsf{SatMC2}$ required in most cases, and in particular for
$k\in\{2,3\}$, up to 10 times more variables than $\mathsf{SatMC1}$
did. For sparse graphs, the number of clauses required by both methods
was similar. The time $\mathsf{SatMC\{1,2\}}$ needed for the generation
of CNF formulas is {\it included} in the running times given below.

\paragraph{Results for real-life graphs.}
Tables~\ref{table:res_dimacs_k2},~\ref{table:res_dimacs_k3},
and~\ref{table:res_dimacs_k4} show the running times (in seconds) and
the optimality gaps for the DIMACS instances solved by $\mathsf{IPMC,
  B\&B, VNS}$, and $\mathsf{SatMC\{1,2\}}$ for $k\in\{2,3,4\}$,
respectively.
\begin{table}
  \centering
  \caption{Computational results of solving M$k$CP for $k=2$ on DIMACS 
    instances.}
  \begin{tabular}{lrp{1pt}rrp{1pt}rrp{1pt}rrp{1pt}rr}
    \toprule
    \multirow{2}{*}{Instance} & 
    \multirow{1}{40pt}{\centering $\mathsf{IPMC}$} & 
    &
    \multicolumn{2}{c}{$\mathsf{B\&B}$} & 
    &
    \multicolumn{2}{c}{$\mathsf{VNS}$} & 
    &
    \multicolumn{2}{c}{$\mathsf{SatMC1}$} &
    &
    \multicolumn{2}{c}{$\mathsf{SatMC2}$}\\
    \cmidrule{2-2}\cmidrule{4-5}\cmidrule{7-8}\cmidrule{10-11}
    \cmidrule{13-14}     
    &time (s) && time (s) &\hspace*{2pt} gap && time (s)&\hspace*{2pt}
    gap && time (s) &\hspace*{2pt} gap&& time (s) &\hspace*{2pt} gap\\
    \midrule
    adjnoun &  0.28 && 0.16 &0 &&0.52 &0 && 0.01&0&&0.02&0\\
    football &  8.91 &&0.74 &0 && 0.76 &0 && 0.02&0&&0.03&0\\
    jazz &  3.98 &&10.1 &0&& 26.8 & 0 && 0.06&0&&0.21&0\\
    celegansm &  39.7 &&26.3 &0&& 29.5&0 && 0.43&0&&2.3&0\\
    email &  $>$3600 &&39.2&0&&39.6 & 0 && 16.1&0&&24.1&0\\
    polblogs &  74.9&&1351&0&&1359 & 0 && 46.9&0&&90.7&0\\
    add20 &  63.9& &126.6&0&&141.6 &0 && 108.2&0&&129.9&0\\
    data &  $>$3600&&$>$3600&0.18&&$>$3600&0.22&&28.8&0&&36.3&0\\
    3elt & $>$3600 &&$>$3600&0.29 &&$>$3600&0.29&&105.9&0&&87.3&0\\
    add32 & $>$3600 &&59.5&0&&136.5 & 0 && 80.1&0&&50.2&0\\
    hep-th & $>$3600 &&199.2 &0&&242.5 &0&&624.1&0&&273.3&0\\     
    whitaker3& $>$3600&&$>$3600&0.36&&$>$3600&0.36&&1091&0&&1045&0\\     
    \bottomrule
  \end{tabular}
  \label{table:res_dimacs_k2}
\end{table}
For $\mathsf{IPMC}$ no optimality gaps are reported since for the
unsolved instances the method could not find any feasible solution, nor
give upper bounds for the solution in the given time limit.
\begin{table}[t]
  \centering
  \caption{Computational results of solving M$k$CP for $k=3$ on DIMACS 
    instances.}
  \begin{tabular}{lrp{1pt}rrp{1pt}rrp{1pt}rrp{1pt}rr}
    \toprule
    \multirow{2}{*}{Instance} & 
    \multirow{1}{40pt}{\centering $\mathsf{IPMC}$} & 
    &
    \multicolumn{2}{c}{$\mathsf{B\&B}$} & 
    &
    \multicolumn{2}{c}{$\mathsf{VNS}$} & 
    &
    \multicolumn{2}{c}{$\mathsf{SatMC1}$}&
    &
    \multicolumn{2}{c}{$\mathsf{SatMC2}$}\\
    \cmidrule{2-2}\cmidrule{4-5}\cmidrule{7-8}\cmidrule{10-11}     
    \cmidrule{13-14}     
    &time (s) && time (s) &\hspace*{2pt} gap && time (s)&\hspace*{2pt}
    gap && time (s) &\hspace*{2pt} gap&& time (s) &\hspace*{2pt} gap\\
    \midrule
    adjnoun &1.74&& 1.99& 0&& 8.25&0&& 0.02&0&&0.19&0\\
    football &86.6&& 12.9& 0&& 12.5&0&& 0.32&0&&0.68&0\\
    jazz &22.4&& 9.8& 0&& 829.9&0 && 1.05&0&&5.21&0\\
    celegansm &46.1&& 155.3& 0&& $>$3600&0.16&& 1.06&0&&26.9&0\\
    email &$>$3600&&$>$3600&0.21&& $>$3600& 0.17&& $>$3600&0.07&&$>$3600&0.07\\
    polblogs&$>$3600&&$>$3600&0.01&&$>$3600&0.01&&$>$3600&0.01&&$>$3600&0\\
    add20 &$>$3600&& $>$3600& 0.02&& $>$3600& 0.02&& 160.8&0&&106.4&0\\
    data &$>$3600&& $>$3600& 0.32&& $>$3600& 0.22&& 72.5&0&&84.1&0\\
    3elt &$>$3600&& $>$3600& 0.46&& $>$3600& 0.33&& 124.1&0&&133.6&0\\
    add32 &$>$3600&&487.5 & 0&& 527.6&0&& 48.5&0&&62.5&0\\
    hep-th &$>$3600&&$>$3600&0.12&&$>$3600&0.12&&925.7&0&&1336&0\\     
    whitaker3 &$>$3600&& $>$3600& 0.46&& $>$3600&0.41&&1372&0&&1413&0\\     
    \bottomrule
  \end{tabular}
  \label{table:res_dimacs_k3}
\end{table}

For $k=2$, $\mathsf{SatMC\{1,2\}}$ were the only methods which solved
optimally all instances within the time limit. In all but three cases
(add20, polblogs, and hep-th), their running times were significantly
better than those of the other methods. Here, $\mathsf{SatMC1}$ was for
medium-size dense instances faster than $\mathsf{SatMC2}$, whereas the
latter was better for large sparse graphs. $\mathsf{B\&B}$ and
$\mathsf{VNS}$ performed similarly solving efficiently small and
medium-size instances.

For $k\in\{3,4\}$, $\mathsf{SatMC1}$, followed by $\mathsf{SatMC2}$, was
the best method regarding the running times, the number of solved
\begin{table}
  \centering
  \caption{Computational results of solving M$k$CP for $k=4$ on DIMACS 
    instances.}
  \begin{tabular}{lrp{1pt}rrp{1pt}rrp{1pt}rrp{1pt}rr}
    \toprule
    \multirow{2}{*}{Instance} & 
    \multirow{1}{40pt}{\centering $\mathsf{IPMC}$} & 
    &
    \multicolumn{2}{c}{$\mathsf{B\&B}$} & 
    &
    \multicolumn{2}{c}{$\mathsf{VNS}$} & 
    &
    \multicolumn{2}{c}{$\mathsf{SatMC1}$}&
    &
    \multicolumn{2}{c}{$\mathsf{SatMC2}$}\\
    \cmidrule{2-2}\cmidrule{4-5}\cmidrule{7-8}\cmidrule{10-11}     
    \cmidrule{13-14}     
    &time (s) && time (s) &\hspace*{2pt} gap && time (s)&\hspace*{2pt}
    gap && time (s) &\hspace*{2pt} gap&& time (s) &\hspace*{2pt} gap\\
    \midrule
    adjnoun &2.18&& 0.86& 0&& 0.94&0&& 0.34&0&&0.49&0\\
    football &177.4&& 0.01& 0&& 0.01&0&& 0.22&0&&0.92&0\\
    jazz &492.4&& 8.87& 0&& 9.11&0 &&137,9 &0&&21.8&0\\
    celegansm &$>$3600&& 186.1& 0&& 194.5&0&& 45.3&0&&95.1&0\\
    email &$>$3600&&$>$3600&0.03&& $>$3600& 1&& $>$3600&0.61&&$>$3600&0.51\\
    polblogs&$>$3600&&$>$3600&1&&$>$3600&1&& $>$3600&0.37&&$>$3600&0.15\\
    add20 &$>$3600&& $>$3600& 1&& $>$3600& 1&&514.8&0&&594.1&0\\
    data &$>$3600&& $>$3600& 0.31&& $>$3600& 0.31&& 112.5&0&&151.4&0\\
    3elt &$>$3600&& $>$3600& 0.54&& $>$3600& 1&& 171.1&0&&186.7&0\\
    add32 &$>$3600&&3322 & 0&& $>$3600&1&& 66.1&0&&71.2&0\\
    hep-th &$>$3600&&$>$3600&1&&$>$3600&1&&$>$3600&0&&$>$3600&0\\     
    whitaker3 &$>$3600&& $>$3600& 0.5&& $>$3600&1&&1563&0&&1687&0\\     
    \bottomrule
  \end{tabular}
  \label{table:res_dimacs_k4}
\end{table}
instances, and the optimality gaps. Only one large (hep-th for $k=4$)
and two medium-size instances email and polblogs could not be solved
optimally by our methods; nevertheless good approximate solutions could
be found. When comparing the two branch-and-bound techniques,
$\mathsf{B\&B}$ solved the same number of instances as $\mathsf{VNS}$,
but was faster. However, the latter delivered for $k=3$ better
approximations. Interestingly, for $\mathsf{SatMC\{1,2\}}$, the running
times required for M3CP on medium-size graphs add20, data, 3elt, and
add32 were longer but still comparable with those needed for M2CP on
those instances. The worst performance across all $k$ tested here showed
$\mathsf{IPMC}$.
\begin{table}[t]
  \centering
  \caption{Computational results of solving M$k$CP for $k=2$ on random 
    instances. The number of unsolved instances is 
    indicated in brackets.}
  \begin{tabular}{ccccrrp{2pt}rrp{2pt}rrp{2pt}rr}
    \toprule
    \multirow{2}{*}{\centering $D$} & 
    \multirow{2}{21pt}{\centering $n$} & 
    \multirow{2}{20pt}{\centering NDV} &
    \multirow{2}{26pt}{\centering $\overline{\omega}_2$} &
    \multicolumn{2}{c}{$\mathsf{IPMC}$} && 
    \multicolumn{2}{c}{$\mathsf{VNS}$} && 
    \multicolumn{2}{c}{$\mathsf{SatMC1}$} &&
    \multicolumn{2}{c}{$\mathsf{SatMC2}$} \\
    \cmidrule{5-6}\cmidrule{8-9}\cmidrule{11-12}\cmidrule{14-15}         
    &&&&time (s) &\hspace*{1pt} gap && time (s) &\hspace*{1pt} gap && 
    time (s)&\hspace*{1pt} gap&&time (s) &\hspace*{1pt} gap\\
    \midrule
    \multirow{6}{*}{0.10}& \multirow{2}{*}{100} &
    min & 21 & 18.4 &0&& 1.66 &0&&0.06 &0 &&0.08&0\\
    & & max & 24.5 & 32.5 &0&& 2.51 &0&&0.07 &0 &&0.09&0\\
    & \multirow{2}{*}{150} &
    min & 26.6 & 3436(9) &0.26&& 41.1 &0&&1.64 &0&&1.84&0\\
    & & max & 32.8 & 3600(10) &0.21&& 61.6 &0&&3.77 &0&&3.84&0\\
    & \multirow{2}{*}{200} &
    min & 33.6 & 3600(10)&1.37&& 1018&0&&30.1 &0 &&31.5&0\\
    & & max &43.1 & 3600(10)&0.74&& 2051(3)&0.05&&204.5 &0 &&176.6 &0\\
    \cmidrule{2-15}
    \multirow{6}{*}{0.15}& \multirow{2}{*}{100} &
    min & 32.3 & 922.1 &0&& 59.4 &0&&1.03 &0&&1.13 &0\\
    & & max & 42.5 & 11.6 &0&& 15.6 &0&&0.25 &0&&0.29 &0\\
    & \multirow{2}{*}{150} &
    min & 53.3 & 3600(10) &0.53&&3600(10) &0.26&&539.8 &0 &&575.2 &0\\
    & & max & 80.8 & 545.9(1) &0.4&& 429 &0&&23.3  &0&&24.2 &0\\
    & \multirow{2}{*}{200} &
    min & 79.3&3600(10)&1.88&&3600(10)&0.43&&3600(10) &0.05&&3600(10) & 0.04\\
    & & max & 124.4 & 107.9&0&& 2005(3)&0.01&&1530(3) &0.01&&1591(3) &0.01\\
    \cmidrule{2-15}
    \multirow{6}{*}{0.20}& \multirow{2}{*}{100} &
    min & 65 & 7.74 &0&& 61.3 &0&&0.99 &0&&1.09 &0\\
    & & max & 68.9 & 0.51 &0&& 11.3 &0&&0.08 &0&&0.16 &0\\
    & \multirow{2}{*}{150} &
    min & 129.8 & 1.48 &0&& 111 &0&&0.73  &0&&1.11 &0\\
    & & max & 122.7 & 1.35 &0&& 50.2 &0&&0.67&0&&0.92 &0\\
    & \multirow{2}{*}{200} &
    min & 192.4 & 3.34&0&& 113&0&&0.13 &0&&0.52 &0\\
    & & max & 176.8&  3.66&0&& 111&0&&0.51 &0&&1.53 &0\\
    \bottomrule
  \end{tabular}
  \label{table:res_random_k2}
\end{table}

\paragraph{Results for random graphs.}
Tables~\ref{table:res_random_k2} and~\ref{table:res_random_k3} present
the results of solving M$k$CP on random graphs. We restrict the results
to $k\in\{2,3\}$. Since $\mathsf{B\&B}$ and $\mathsf{VNS}$ performed on
random graphs tested similarly, we provide only the results of
$\mathsf{VNS}$, the one of a better overall performance. Average running
times and average optimality gaps, both computed across the 10 graph
samples of a given $D$, $n$, and NDV, are reported. Additionally, for
each instance category, i.e., for a given $D$, $n$, and NDV, we provide
average 2- and 3-club numbers $\overline{\omega}_2$ and
$\overline{\omega}_3$, computed for each category from the 10 (optimum)
values of $\omega_2$ and $\omega_3$ found by our methods. Noteworthy, to
compute $\omega_2$ and $\omega_3$, only for 13 of a total of 360 graph
samples more than one hour was needed. Finally, the average optimality
gap for $\mathsf{IPMC}$ was calculated from the gap values returned by
the integer routine of CPLEX.

For $k=2$ and average densities $D=0.10$ and $0.20$, $\mathsf{SatMC\{1,
  2\}}$ found optimal solutions for every instance, whereas the running
times were by far shorter than those of the other methods. For graphs of
$D=0.15$, being reportedly the hardest
ones~\cite{pajouh2012oninclusionwise}, our methods obtained optimal
solutions except for 13 test samples of size $n=200$, for which,
however, competitive approximate solutions could be given. When
comparing the performance for densities $>0.10$, $\mathsf{SatMC\{1,2\}}$
solved instances with maximum NDV always faster than those of the same
size and density but with minimum NDV. This could not be observed for
the other methods. However, for $D=0.15$ and all methods, the instances
with minimum NDV turned out to be much harder than those with maximum
NDV. $\mathsf{IPMC}$ was the slowest method regarding graphs of $D=0.10$
and $0.15$, but it performed exceptionally well for $D=0.20$, taking the
third place by beating $\mathsf{VNS}$.

For $k=3$, our methods were able to solve all instances into optimality,
whereas the average running times required were considerably shorter
than those of the other methods. This held also for instances of
challenging densities $0.035$ and $0.05$ according
to~\cite{almeida2012integer,bourjolly2002exact}. $\mathsf{VNS}$
exhibited the third best performance, solving optimally all but 12
instances. However, for $D=0.10$, it was beaten by $\mathsf{IPMC}$. For
both values of $k$, $\mathsf{SatMC1}$ performed slightly better than
$\mathsf{SatMC2}$, primarily due to smaller size of the CNF encodings.

Finally, for a given $k, n$, and NDV, as the density $D$ increased, the
average solution size found by all methods increased, too, while the
\begin{table}[t]
  \centering
  \caption{Computational results of solving M$k$CP for $k=3$ on random 
    instances. The number of unsolved instances is 
    indicated in brackets.}
  \begin{tabular}{ccccrrp{2pt}rrp{2pt}rrp{2pt}rr}
    \toprule
    \multirow{2}{*}{\centering $D$} & 
    \multirow{2}{21pt}{\centering $n$} & 
    \multirow{2}{20pt}{\centering NDV} &
    \multirow{2}{26pt}{\centering $\overline{\omega}_3$} &
    \multicolumn{2}{c}{$\mathsf{IPMC}$} && 
    \multicolumn{2}{c}{$\mathsf{VNS}$} && 
    \multicolumn{2}{c}{$\mathsf{SatMC1}$} &&
    \multicolumn{2}{c}{$\mathsf{SatMC2}$}\\
    \cmidrule{5-6}\cmidrule{8-9}\cmidrule{11-12}\cmidrule{14-15}         
    &&&&time (s) &\hspace*{2pt}gap && time (s) &\hspace*{2pt}gap && 
    time (s)&\hspace*{2pt}gap&&time (s) &\hspace*{2pt}gap\\
    \midrule
    \multirow{6}{*}{0.035}& \multirow{2}{*}{100} &
    min & 24 & 1.91&0 && 1.13& 0&&0.02 &0 &&0.05 &0 \\
    & & max & 27.1& 1.94&0 && 1.31& 0&&0.01 &0 &&0.06 &0 \\
    & \multirow{2}{*}{150} &
    min & 31.6& 95.3& 0&& 7.49& 0&&0.18 &0 &&0.41 &0 \\
    & & max & 33.7& 213.6&0 && 10.4&0 &&0.23 &0 &&0.49 &0 \\
    & \multirow{2}{*}{200} &
    min & 36.9& 2381(4)& 0.12&& 54.2&0 &&1.54 & 0&&2.91 & 0\\
    & & max & 40.8& 3105(7)& 0.18&& 107.8&0 &&2.86 & 0&&5.08 & 0\\
    \cmidrule{2-15}
    \multirow{6}{*}{0.05}& \multirow{2}{*}{100} &
    min & 30.4& 7.69& 0&& 3.25&0 &&0.04 & 0&&0.11 & 0\\
    & & max & 33.9& 6.14& 0&& 3.18& 0&&0.04 &0 &&0.14 & 0\\
    & \multirow{2}{*}{150} &
    min & 43.9&2750(6) &0.06 &&130.2 &0 &&1.71 &0 &&2.96 & 0\\
    & & max &56 &1138(2) &0.02 &&128.1 &0 &&2.05 &0 &&4.04 & 0\\
    & \multirow{2}{*}{200} &
    min & 55& 3600(10)&0.77 &&3578(9) & 0.19&&91.4 & 0 &&128.7 & 0\\
    & & max & 84.3& 3168(8)&0.21&&2311(3)&0.06&&65.9 &0 &&117.1 & 0\\
    \cmidrule{2-15}
    \multirow{6}{*}{0.10}& \multirow{2}{*}{100} &
    min & 82.8& 1.19&0 && 8.77& 0&&0.03 & 0&&0.31 & 0\\
    & & max & 81.8& 1.36& 0&& 8.86& 0&&0.03 &0 &&0.31 & 0\\
    & \multirow{2}{*}{150} &
    min & 146& 10.1& 0&& 42.8& 0&&0.12 &0 &&1.51 & 0\\
    & & max & 141.5& 13.2&0 && 46.8& 0&&0.12 & 0&&1.46 & 0\\
    & \multirow{2}{*}{200} &
    min & 199.2& 11.9& 0&& 105.4& 0&&0.33 & 0&&5.88 & 0\\
    & & max & 196.5& 13.4&0 && 197.2&0 &&0.51 & 0&&6.32 & 0\\
    \bottomrule
  \end{tabular}
  \label{table:res_random_k3}
\end{table}
average running time increased first up to a peak and then declined for
higher densities. This can be explained by the fact that sparse graphs
are easier to solve, because there are less possibilities to construct a
$k$-club, whereas for higher densities many of the problems are becoming
trivial. More specifically, for $D\geq 0.20$ and $k=2$, and for $D\geq
0.10$ and $k=3$, the values of $\overline{\omega}_k$ approached $n$ and
the instances became easier to solve despite their growing sizes. The
peak average running time can be used to determine the challenging
densities for an algorithm for M$k$CP~\cite{pajouh2012oninclusionwise}.
\mbox{Table}~\ref{table:densities} gives those densities identified
empirically for all methods tested. The numerical results show that for
a given $k$, the challenging densities were the same for minimum and
maximum NDV instances, and decreased as $n$ increased. This effect was
most evident for $\mathsf{IPMC}$, followed by $\mathsf{B\&B}$ and
$\mathsf{VNS}$. $\mathsf{SatMC\{1,2\}}$ turned out to be least affected,
indicating clearly their better robustness.
\begin{table}[t]
  \centering
  \caption{Challenging densities $D$ for solving M$k$CP on random 
    instances.}
  \begin{tabular}{ccccp{2pt}ccc}
    \toprule
    \multirow{2}{45pt}{\centering Method}  
    &\multicolumn{3}{c}{$k=2$} && \multicolumn{3}{c}{$k=3$}\\
    \cmidrule{2-4}\cmidrule{6-8}
    &$n=100$\hspace*{8pt}&$n=150$\hspace*{8pt}&$n=200$&
    &$n=100$\hspace*{8pt}&$n=150$\hspace*{8pt}&$n=200$\\
    \midrule
    $\mathsf{IPMC}$ & $0.15$ & $0.1, 0.15$ & 0.1, 0.15 && 0.05& 0.05&
    0.035, 0.05\\
    $\mathsf{VNS}$ & 0.15, 0.2 & 0.15 & 0.1, 0.15 && 0.1& 0.05& 0.05\\
    $\mathsf{SatMC\{1,2\}}$ & 0.15 & 0.15 & 0.15 && 0.05& 0.05& 0.05\\
    \bottomrule
  \end{tabular}
  \label{table:densities}
\end{table}

\section{Conclusion}
\label{s:conclusion}
In this paper, we presented two PARTIAL MAX-SAT formulations of M$k$CP
for a positive integer $k$. Using those encodings, we implemented two
exact methods for M$k$CP, $\mathsf{SatMC1}$ and $\mathsf{SatMC2}$, and
evaluated them experimentally for typical values of $k\in\{2,3,4\}$ both
on real-life as well as on random graphs. The computational study showed
that our approach outperforms other state-of-the-art algorithms
developed in the last years. It computed optimal solutions in most cases
much faster and found good approximate solutions in case the computation
had to be terminated. Its short running times on small and moderate-size
instances qualify it clearly for usage in interactive tools, e.g., for
clustering biological networks~\cite{balasundaram2005novel}, providing
useful insights into substructures in those networks.

It would be of interest to adapt our ideas for solving other clique
relaxations like $k$-clique, $k$-plex, or $R$-robust
$k$-club~\cite{balasundaram2011clique,seidman1978graph,veremyev2012identifying}. Moreover,
for $k=2$, our approach could also be compared with the parameterized
algorithm of Hartung et al.~\cite{hartung2012parametrized}, and for a
general $k$ with the branch-and-bound method by Chang et
al.~\cite{chang2013finding}. One could also evaluate our methods for
solving M$k$CP for $k>4$ on power-law graphs from bioinformatics and
social web applications. Finally, it is an open question, if any of the
algorithmic ideas of modern CDCL SAT-solving, from which our approach
clearly benefits, could successfully be extended to BIP.

\subsubsection*{Acknowledgments.}
The author would like to thank Shahram Shahinpour and Sergiy Butenko for
providing an implementation of their method
from~\cite{shahinpour2013algorithms}.

\bibliographystyle{splncs03}
\bibliography{SATliterature}

\end{document}